\documentclass[twocolumn,apl,showpacs,preprintnumbers,amsmath,amssymb]{revtex4}
\usepackage{graphicx}
\usepackage{dcolumn}
\usepackage{bm}
\usepackage{color}

\begin{document}

\title{Origin of the emission within the cavity mode of coupled quantum dot-cavity systems}

\author{J. Suf\mbox{}fczy\'nski, A. Dousse, K. Gauthron, A. Lema\^itre, I.
Sagnes, L. Lanco, P. Voisin, J. Bloch and P. Senellart}
\affiliation{Laboratoire de Photonique et Nanostructures, LPN/CNRS,
Route de Nozay, 91460 Marcoussis, France}

\date{\today}

\begin{abstract}
The origin of the emission within the optical mode of a coupled
quantum dot-micropillar system is investigated. Time-resolved
photoluminescence is performed on a large number of
deterministically coupled devices in a wide range of temperature and
detuning. The emission within the cavity mode is found to exhibit
the same dynamics as the spectrally closest quantum dot state. Our
observations indicate that fast dephasing of the quantum dot state
is responsible for the emission within the cavity mode. An
explanation for recent photon correlation measurements reported on
similar systems is proposed.
\end{abstract}

\maketitle

An individual semiconductor Quantum Dot (QD) coupled to an optical
cavity mode is a promising system in view of practical
implementation of efficient single photons
sources~\cite{Santori2002}, nanolasers~\cite{Strauf2006} as well as
remote quantum bit entanglement \cite{Rarity08}. Early experiments
 of solid state cavity quantum electrodynamics (CQED)
relied on the random possibility of finding a quantum dot spectrally
and spatially resonant with a single cavity mode
\cite{Reithmaier04,Yoshie04,Peter05}. In these first demonstrations,
the yield of the fabrication process was very small and hindered the
possibility of performing systematic investigations. Recently, many
technological advances like deterministic QD-cavity
coupling~\cite{Badolato05,Dousse08}, tuning techniques
\cite{Strauf06,FinleyArx08} electrical pumping of the
devices~\cite{Bockler08} have enabled a better control of the
system. As a result, the number of optical studies of solid state
CQED has increased~\cite{Henessy07,Press07,KaniberPRB08, Balet07,
Srinivasan08} and it now appears that the system cannot simply be
described in the framework of the Jaynes-Cumming model as for CQED
with atoms. Indeed, when studying the emission of coupled QD-cavity
devices, unexpected questions arise, whether the system is in the
weak coupling regime (Purcell effect)\cite{KaniberPRB08} or strong
coupling regime \cite{Henessy07,Press07}. Why, in the situation
where the spectral detuning between the quantum dot and the optical
mode is much larger than the mode linewidth, emission is still
mostly observed at the mode energy? Why the emission within the
cavity mode is anticorrelated with the QD emission even for
detunings as large as ten mode
linewidths~\cite{Henessy07,Press07,KaniberPRB08}?

In the present work, we study the emission dynamics of coupled
QD-pillar cavities in the weak coupling regime. Each pillar embeds a
single quantum dot deterministically coupled to its fundamental
optical mode. Time resolved photoluminescence measurements are
carried out in a quantum dot-mode detunings range extending up to
three mode linewidths. The emission within the cavity mode is found
to present the same dynamics as the spectrally closest QD state, in
a wide temperature range and independently of the detuning sign.
This is observed whether the decay of the QD state emission is given
by its radiative lifetime or is driven by scattering processes. When
two states of the same QD are equally detuned from the cavity mode,
the emission within the mode presents the same decay as the
spectrally wider QD state. These observations indicate that the
emission within the cavity mode is driven by fast dephasing
processes \cite{Cassabois06,Peter04}, in accordance with recent
theoretical proposals \cite{Naesby08, Auff09}. Our results also
provide an overall explanation for recent photon correlation
measurements reported on similar systems
\cite{Henessy07,KaniberPRB08}.

Pillar microcavities were fabricated using a planar microcavity
constituted by 20 (24) pairs of
Al$_{0.1}$Ga$_{0.9}$As/Al$_{0.95}$Ga$_{0.05}$As Bragg mirrors. The
cavity exhibits a quality factor of 4500 and embeds a layer of
self-organized InAs/GaAs QDs. Rapid thermal annealing of the sample
(30 s at 850${^\circ}$) allows reducing the spatial density of QDs
emitting at the planar cavity mode energy to less than 1
$\mu$m$^{-2}$ by blueshifting the overall quantum dot
distribution~\cite{Langbein04}. Micropillars containing an
individual QD spectrally and spatially matched to the microcavity
fundamental mode are fabricated with the single-step in-situ
lithography technique detailed in Ref.~\onlinecite{Dousse08}. Even
though higher energy QDs are also embedded in the pillar, the
deterministic coupling between a single low energy QD and the pillar
fundamental mode allows neglecting the influence of other QD in a
wide range of detunings. Pillar diameters range from 1.5 $\mu$m to
2.3 $\mu$m. The sample is placed in a cold finger cryostat. The
Photoluminescence (PL) is excited non-resonantly at 1.59 eV with a
pulsed Ti:Sapphire laser. The same microscope objective is used to
focus the laser beam on the pillars and to collect the emission
which is sent to a spectrometer. The emission is then detected by a
CCD camera with a 150 $\mu$eV spectral resolution or by a Streak
Camera for time resolved analysis with a resolution of 20 ps. 
\begin{figure}
{\includegraphics[width=85mm]{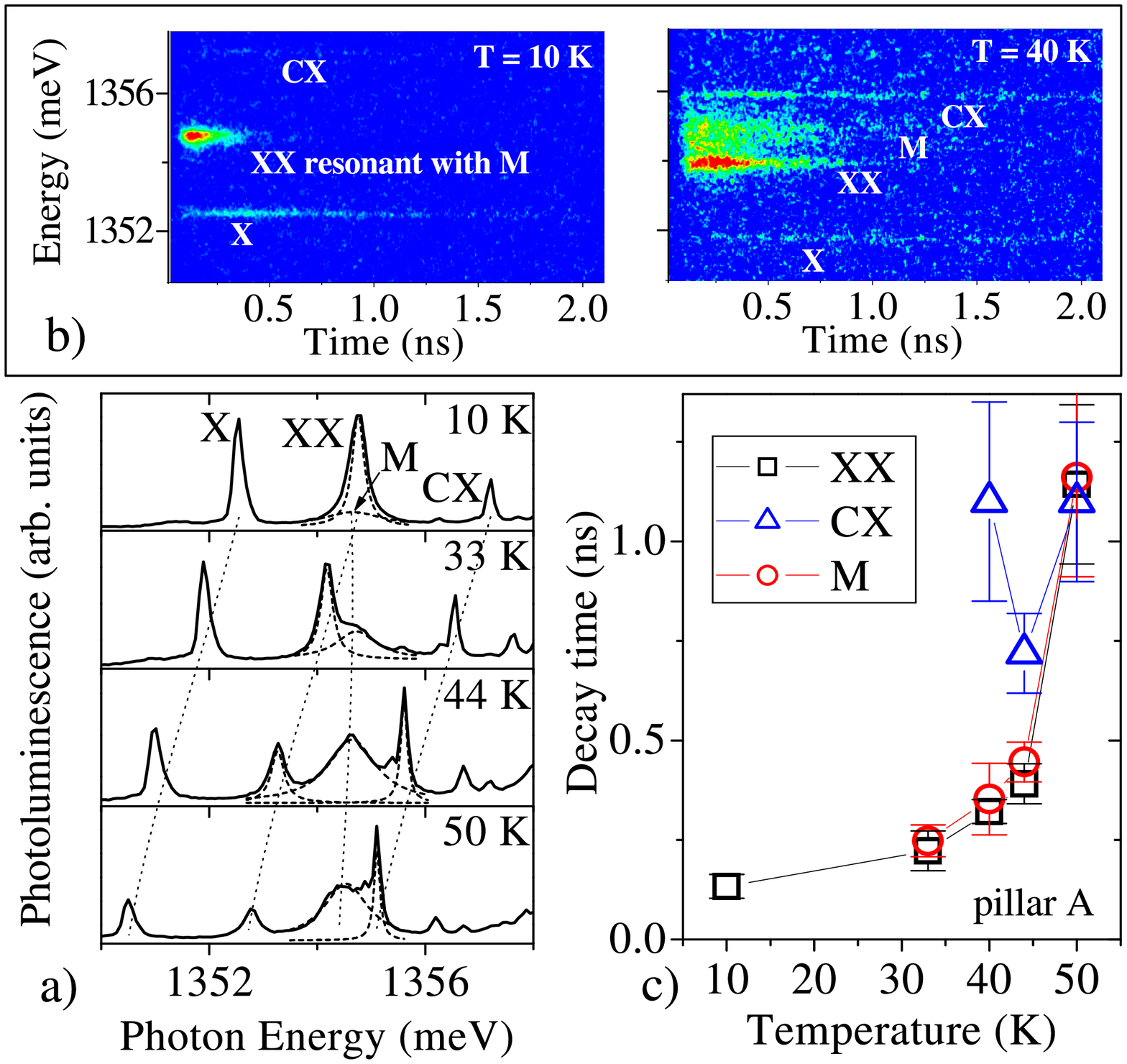}}
\caption{\label{fig:pillarA}(Color online) (a) Time integrated
emission spectra of pillar A taken at temperatures $T$ between 10 K
and 50 K. Excitation power 0.25 $\mu$W.  Dotted curves help
following the spectral position of the X, CX, XX and M lines with
temperature. (b) Streak camera images (emission intensity as a
function of energy and time) measured  on pillar A at $T$ = 10~K and
$T$ = 40~K. (c) Emission decay time for XX, CX and M lines as a
function of $T$ between 10 K and 50 K.}
\end{figure}

Time resolved photoluminescence is measured on seven pillars. The
detuning $\Delta$ between the cavity mode and the QD state is tuned
over a wide range by varying the temperature. Change of the
temperature modifies not only the magnitude of the Purcell effect on
the QD state, but also carrier scattering toward the QD higher
energy states. As a result, various and complex dependences can be
observed when studying the emission dynamics of the QD emission line
as a function of temperature. We investigate the emission dynamics
within the cavity mode for these various situations and start with
illustrating of our findings by presenting the measurements on two
pillars A and B.

Pillar A presents a 1.7 $\mu$m diameter and a quality factor Q=1300.
It embeds a QD in its center experiencing a Purcell factor of
$F_P$=7. Figure~\ref{fig:pillarA}(a) shows the time integrated PL
spectra from pillar A taken in the 10~K to 50~K  temperature range.
The emission lines are identified through power and temperature
dependent measurements as coming from recombination of neutral
exciton (X), biexciton (XX) and charged exciton (CX) confined in the
same QD, as well as emission within the cavity mode (M). At 10 K,
the fundamental mode M is resonant to XX. The streak camera image
recorded at $T$ = 10 K is shown in Figure~\ref{fig:pillarA}(b). The
radiative cascade between the XX and X state is seen as a delay of
the X signal rise with respect to XX signal. The XX decay time is
$\tau_{XX}$ = 135 $\pm$ 30 ps, much shorter than the typical 1 ns
decay time in these QDs, showing the strong Purcell effect
experienced by the XX. When increasing temperature, the XX-M
detuning as well as the XX decay time increase. Measurements
performed on QD emission lines well detuned from the cavity mode
show no change of the decay times between 10 K and 45 K. As a
result, the increase if the XX decay time for pillar A can fully be
attributed to the gradual quenching of the Purcell effect below 45
K.

When the XX-M detuning is large enough, the decay time of the
emission within the cavity mode is  extracted from the streak camera
images. The measured decay time of the emission within the mode is
plotted as a function of temperature together with the XX decay time
in Fig.~\ref{fig:pillarA}(c). Remarkably, in the whole 10K - 45 K
temperature range, both XX and M lines present the same decay time.
The same behavior is observed on each pillar for which the emission
dynamics of the QD state is determined by its radiative decay. We
now show that the same behavior is also observed when the QD state
dynamics is no longer governed by its radiative decay.

Indeed, around 50 K, scattering of carriers out off the QD
fundamental state leads to an increase of the decay time of the QD
states \cite{BQtemperature}: confined carriers are scattered to  the
s-p non-radiative states and the decay time of the QD line is no
longer determined by its radiative lifetime. Influence of these
scattering processes is seen on the CX line which decay shortens due
to Purcell effect between 40 K and 44 K, but increases at 50 K even
though CX is closer to resonance with the cavity mode (Fig.
~\ref{fig:pillarA}(c)).
\begin{figure}
\includegraphics[width=70mm]{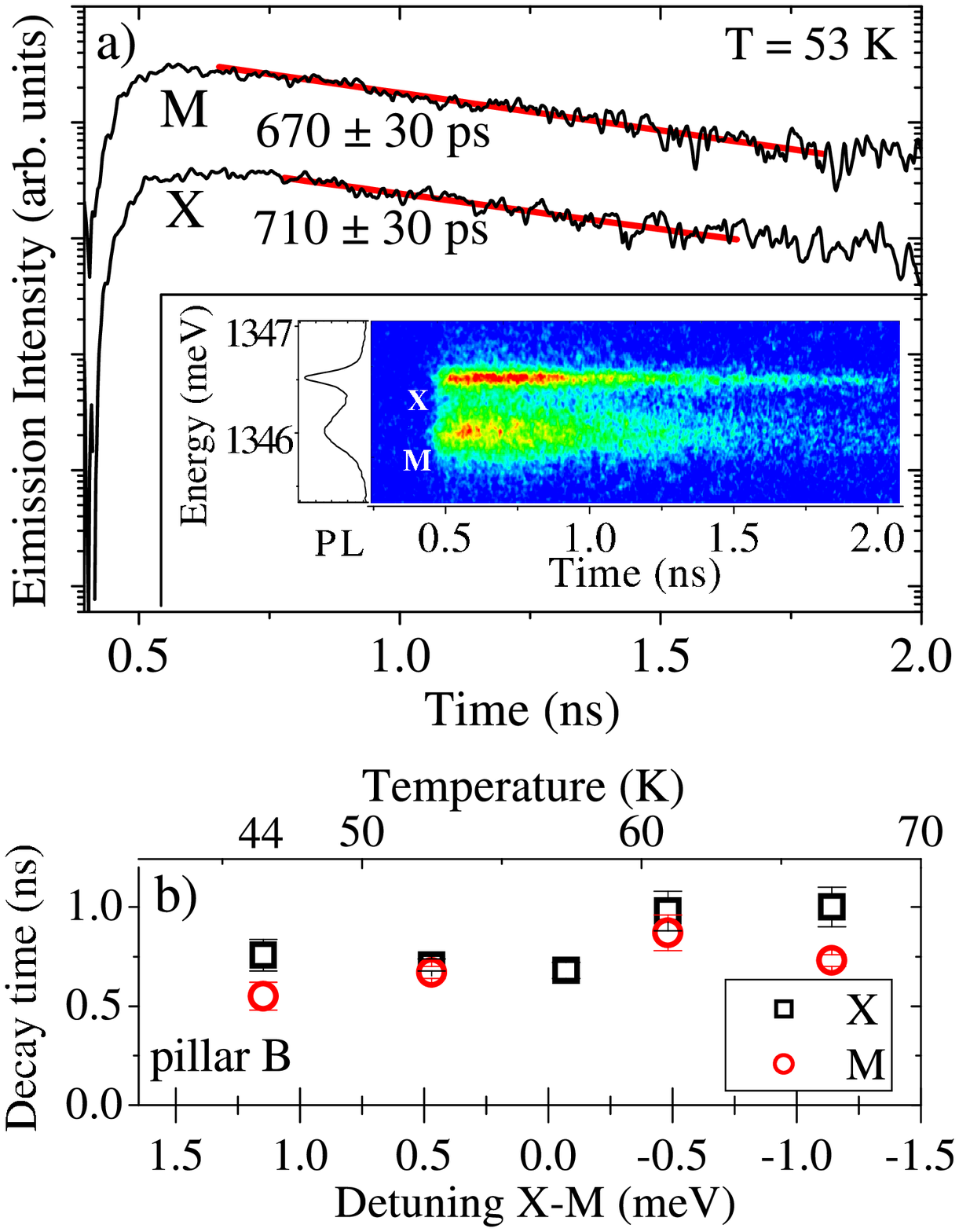}
\caption{\label{fig:pillarB1}(Color online) (a) Emission intensity
as a function of time at X and M wavelengths measured on pillar~B at
$T$ = 53 K (log. scale).  The curves have been vertically shifted
for clarity. Decay times obtained from a monoexponential fit are
indicated. Inset: corresponding streak camera image, as well as time
integrated spectrum.(b) Emission decay time at X and M wavelengths
plotted versus X-M detuning (bottom axis) and temperature (top axis)
for pillar B.}
\end{figure}

\begin{figure}
\includegraphics[width=70mm]{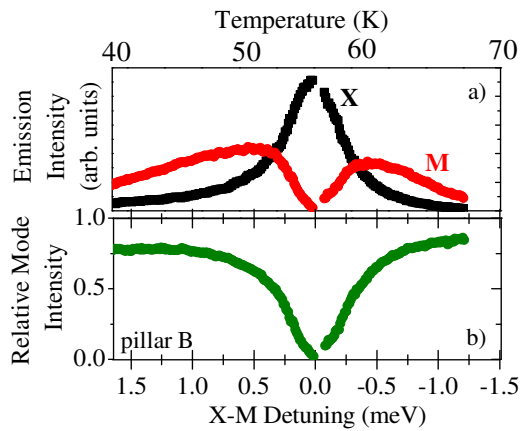}
\caption{\label{fig:pillarB2}(Color online) (a) Intensity of X and M
emission lines measured on pillar B. (b) Ratio of emission intensity
within the mode to the overall  (X plus M) emission intensity. The
plots are versus X-M detuning (bottom axis) and temperature $T$ (top
axis).}
\end{figure}

To illustrate the properties of the mode emission at elevated
temperature, we present the measurements performed on pillar B
(diameter 2.3 $\mu$m, $\gamma_{M}$ = 0.45 meV, Q=3000, $F_P=$8) for
which the X-M resonance occurs around 57 K.
Figure~\ref{fig:pillarB1}(b) shows the X decay time as a function of
temperature and X-M detuning. Although the signal from X state is
greatly enhanced due to Purcell effect (see Fig.
~\ref{fig:pillarB2}(a)) when X is tuned to resonance with M, no
shortening of the X emission decay is observed. This is because the
observed X emission decay time is governed by the
 lifetime of the carriers in higher energy states above 50 K.
However, as for pillar A, in the whole temperature range, the decay
time of the emission within the cavity mode is very close to the
decay time of the excitonic line

\begin{figure}
\includegraphics[width=80mm]{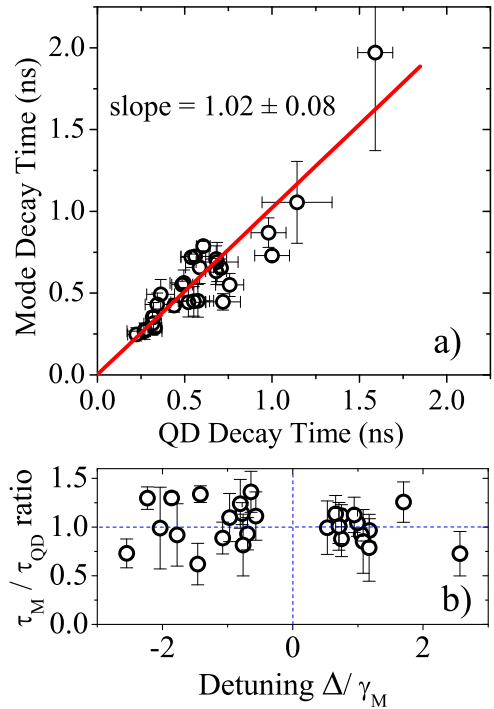}
\caption{\label{fig:Summary}(Color online) (a) Decay time of the
emission within cavity mode $\tau_{M}$ as a function of the emission
decay time for the spectrally closest QD state $\tau_{QD}$ measured
on seven different devices for more than thirty temperatures. Free
parameter linear fit yielding the 1.02 $\pm$ 0.08 slope is shown.
(b) The ratio r = $\tau_{M}$/$\tau_{QD}$ as a function of  QD-M
detuning $\Delta$ normalized by mode linewidth $\gamma_{M}$ of each
pillar.}
\end{figure}

The strong correlation between the decay time within the cavity mode
and the QD emission line observed on both, A and B, pillars is
further investigated on five other pillars, presenting Purcell
factor between 7 and 14, in a range of detunings up  to 3 mode
linewitdhs and for temperatures between 10 K and 70 K. Figure
~\ref{fig:Summary} shows the decay time of the emission within the
cavity mode versus the decay time of the QD state closest in energy,
in collecting the results on all seven pillars. The experimental
points are remarkably located on the diagonal showing that the
emission within the cavity mode arises from the spectrally closest
QD state. This observation is further confirmed by the observation
of no delay between the rise of emission at QD state energy and at
the M energy (Figs.~\ref{fig:pillarA}(b) and
~\ref{fig:pillarB1}(a)). Figure~\ref{fig:Summary}(b) shows the ratio
r = $\tau_{M}$/$\tau_{QD}$ plotted versus the QD-M detuning
normalized by the mode linewidth $\gamma_{M}$ for each micropillar.
Experimental points are scattered around unity with no dependence on
the  detuning $\Delta$ value or sign.

The present observations as well as photon correlation measurements
performed on coupled QD-microcavity mode systems \cite{Henessy07}
indicate that a single quantum emitter is responsible for the
emission at both QD and mode energies. In the framework of
Jaynes-Cumming hamiltonian, the emission is mainly peaked at the QD
energy whatever the QD-M detuning is. Emission at the energy of the
cavity mode is essentially negligible outside the QD-M crossing
region. In a recent theoretical work, Naesby and coworkers
\cite{Naesby08} have investigated the issue of unexpectedly strong
emission within the cavity mode in the case the strong coupling
regime. They show that when pure dephasing broadens the exciton
line, the emission at the energy of the cavity mode is greatly
enhanced. In this framework, the emission dynamics is expected to be
the same both at exciton and mode energies, as observed here. For a
strong dephasing of the QD state, the emission is expected to mainly
take place at the mode energy, independently of the detuning
$\Delta$ in the large detuning range. The fraction of emission
within M is calculated to decrease close to QD-M resonance.

Figure~\ref{fig:pillarB2}(a) plots the emission intensity within the
mode as a function of detuning with the X line, evidencing a strong
decrease close to resonance. This behavior is observed for each
resonance  X, XX or CX with the cavity mode, further confirming the
major role played by fast dephasing. The ratio of the emission
intensity within the cavity mode to overall (X and M) emission
intensity is plotted in Figure~\ref{fig:pillarB2}(b) showing that 80
$\%$ of emission takes place within the mode for large detunings.
Close to resonance, this ratio decreases and goes down to $0.01$.
The reduction of relative mode intensity at X-M resonance is much
stronger in our experiment than calculated in
Ref.~\onlinecite{Naesby08}. Because we change the detuning by
increasing the temperature, one cannot straightforwardly compare our
measurements with the theoretical predictions since not only the
detuning but also the dephasing of the QD state is
modified~\cite{Auff09}.  Moreover, Naesby and coworkers treat the
case of strong coupling for which the emission intensity within each
line is equal at resonance, so that the ratio goes to 0.5
independently of dephasing effects. Further theoretical
investigations would be valuable to quantitatively describe the
influence of fast dephasing in the case of weak coupling regime.

As one would expect, the observation of similar decay times for the
emission from a QD state and the emission within the cavity mode
falls when the detuning is very large. As seen in
Fig.~\ref{fig:pillarA}(a) at 40 K, the cavity mode presents the same
decay as the XX line, closest in energy, while the decay at the
strongly detuned exciton energy is longer. In the framework of QD
emission within the mode mediated by dephasing processes, we note
that when the cavity mode lies between two transitions of the
coupled QD, the emission within the cavity mode is fed by a
combination of the two states. This effect is clearly seen on pillar
A at 44 K. The CX is slightly spectrally closer to the mode than the
XX state at this temperature, yet the decay of the emission within
the mode is mostly driven by the XX decay. Indeed, as can be seen on
the spectra in Fig.~\ref{fig:pillarA}(a), the XX presents a much
larger linewidth (290 $\mu$eV) than the CX ($<150 \mu$eV): this
difference in dephasing results in a stronger contribution of XX to
the emission within the mode. In Fig. ~\ref{fig:Summary}(a), the
decay time at M energy is plotted as a function of the spectrally
closest QD state decay time  with no consideration of the other QD
states and their emission linewidths: most of the off diagonal
experimental points reflect measurements where the mode lies between
two QD lines presenting different dephasing. Finally, when the
cavity mode is strongly detuned from any quantum dot state, we
accordingly observe that the emission at the mode energy presents a
decay time different from any particular QD line as reported in
Ref.~\onlinecite{Henessy07}.

The above  considerations allow us to propose an explanation for the
photon correlation experiments reported in [\onlinecite{Henessy07}]
where some apparent contradiction was seen. While a strong
anticorrelation between the X emission and the emission within the
cavity mode was observed suggesting that both emissions arised from
the same quantum emitter, auto-correlation measurements performed on
the emission within the cavity mode revealed almost no anti-bunching
\cite{Henessy07}. In the large detuning range where these
experiments have been performed, the emission within the cavity mode
results from the sum of various contributions corresponding to fast
dephasing of several states of the same quantum dot. In this
framework, cross correlation between M and  X is a sum of
correlations between X, CX, XX emissions and X emission, weighted by
respective value of detuning, dephasing and intensity of each line
of the QD. In continuous wave excitation, all the correlation
functions contributing to this sum (auto correlation X-X and cross
correlations CX-X and XX-X) reveal an anticorrelation at zero delay
time \cite{Moreau2001} and for negative delays, so that their sum
also exhibits g$^{(2)}$($\tau$=0) $\sim$ 0 and anticorrelation for
negative delay. On the contrary, for positive delay, X-X
autocorrelation exhibits antibunching and CX-X and XX-X cross
correlations bunching which explains the strong asymmetry in the X-M
correlation reported in Ref.~\onlinecite{Henessy07}. The observation
of almost no antibunching in  M-M autocorrelation under pulsed
excitation is explained in a similar way. M-M autocorrelation is the
sum of all possible autocorrelations and cross correlations between
all the states of the same QD. In pulsed excitation measurements,
each cross correlation within the radiative cascade contributes to
the zero delay peak of the correlation histogram and decreases the
level of antibunching.

In conclusion, emission dynamics of deterministically coupled
quantum dot-pillar microcavity systems bring strong evidence that
the emission within the cavity mode results from fast dephasing of
the QD state, in agreement with recent theoretical
works~\cite{Naesby08,Auff09}. Specifically, emission within the
cavity mode is driven by the closest state of the QD, as long as
fast dephasing is similar for each state. When two states are
similarly detuned from the cavity mode, the emission at the mode
energy is driven by the spectrally widest quantum state. We note
that thanks to deterministic coupling, the influence of other QDs
embedded in the pillar at much higher energy is negligible.
Moreover, the fabricated devices enable unambiguous investigations
without the need for a selective QD excitation. Our measurements
provide an overall understanding of the emission mechanism in solid
state CQED systems.
\begin{acknowledgments}
This work was partially supported by the French ANR JCJC MICADOS.
The authors acknowledge stimulating discussions with Pr. A. Imamoglu
and A. Beveratos.
\end{acknowledgments}


\end{document}